\documentclass[final,5p,times,twocolumn]{elsarticle}
\pdfoutput=1
\usepackage{amsmath,amssymb,amsfonts}
\usepackage{graphicx,color,float}
\usepackage{subfigure}
\usepackage{multirow,booktabs}
\usepackage[mathlines]{lineno}

\begin{document}

\begin{frontmatter}

\title{Measurement of the $\omega \to \pi^+ \pi^- \pi^0$ Dalitz plot distribution }

\author[IKPUU]{The WASA-at-COSY Collaboration\\[2ex] P.~Adlarson\fnref{fnmz}}
\author[ASWarsN]{W.~Augustyniak}
\author[IPJ]{W.~Bardan}
\author[Edinb]{M.~Bashkanov}
\author[MS]{F.S.~Bergmann}
\author[ASWarsH]{M.~Ber{\l}owski}
\author[IITB]{H.~Bhatt}
\author[Budker,Novosib]{A.~Bondar}
\author[IKPJ]{M.~B\"uscher\fnref{fnpgi,fndus}}
\author[IKPUU]{H.~Cal\'{e}n}
\author[IFJ]{I.~Ciepa{\l}}
\author[PITue,Kepler]{H.~Clement}
\author[IPJ]{E.~Czerwi{\'n}ski}
\author[MS]{K.~Demmich}
\author[IKPJ]{R.~Engels}
\author[ZELJ]{A.~Erven}
\author[ZELJ]{W.~Erven}
\author[Erl]{W.~Eyrich}
\author[IKPJ,ITEP]{P.~Fedorets}
\author[Giess]{K.~F\"ohl}
\author[IKPUU]{K.~Fransson}
\author[IKPJ]{F.~Goldenbaum}
\author[IKPJ,IITI]{A.~Goswami}
\author[IKPJ,HepGat]{K.~Grigoryev\fnref{fnac}}
\author[IKPUU]{C.--O.~Gullstr\"om}
\author[IKPUU]{L.~Heijkenskj\"old\corref{coau}}\ead{lena.heijkenskjold@physics.uu.se}
\author[IKPJ]{V.~Hejny}
\author[MS]{N.~H\"usken}
\author[IPJ]{L.~Jarczyk}
\author[IKPUU]{T.~Johansson}
\author[IPJ]{B.~Kamys}
\author[ZELJ]{G.~Kemmerling\fnref{fnjcns}}
\author[IKPJ]{F.A.~Khan}
\author[IPJ]{G.~Khatri\fnref{fnharv}}
\author[MS]{A.~Khoukaz}
\author[IPJ]{O.~Khreptak}
\author[HiJINR]{D.A.~Kirillov}
\author[IPJ]{S.~Kistryn}
\author[ZELJ]{H.~Kleines\fnref{fnjcns}}
\author[Katow]{B.~K{\l}os}
\author[IPJ]{W.~Krzemie{\'n}}
\author[IFJ]{P.~Kulessa}
\author[IKPUU,ASWarsH]{A.~Kup{\'s}{\'c}}
\author[Budker,Novosib]{A.~Kuzmin}
\author[NITJ]{K.~Lalwani}
\author[IKPJ]{D.~Lersch}
\author[IKPJ]{B.~Lorentz}
\author[IPJ]{A.~Magiera}
\author[IKPJ,JARA]{R.~Maier}
\author[IKPUU]{P.~Marciniewski}
\author[ASWarsN]{B.~Maria{\'n}ski}
\author[ASWarsN]{H.--P.~Morsch}
\author[IPJ]{P.~Moskal}
\author[IKPJ]{H.~Ohm}
\author[PITue,Kepler]{E.~Perez del Rio\fnref{fnlnf}}
\author[HiJINR]{N.M.~Piskunov}
\author[IKPJ]{D.~Prasuhn}
\author[IKPUU,ASWarsH]{D.~Pszczel}
\author[IFJ]{K.~Pysz}
\author[IKPUU,IPJ]{A.~Pyszniak}
\author[IKPJ,JARA,Bochum]{J.~Ritman}
\author[IITI]{A.~Roy}
\author[IPJ]{Z.~Rudy}
\author[IPJ]{O.~Rundel}
\author[IITB,IKPJ]{S.~Sawant\corref{coau}}\ead{siddhesh.sawant@iitb.ac.in}
\author[IKPJ]{S.~Schadmand}
\author[IPJ]{I.~Sch\"atti--Ozerianska}
\author[IKPJ]{T.~Sefzick}
\author[IKPJ]{V.~Serdyuk}
\author[Budker,Novosib]{B.~Shwartz}
\author[MS]{K.~Sitterberg}
\author[PITue,Kepler,Tomsk]{T.~Skorodko}
\author[IPJ]{M.~Skurzok}
\author[IPJ]{J.~Smyrski}
\author[ITEP]{V.~Sopov}
\author[IKPJ]{R.~Stassen}
\author[ASWarsH]{J.~Stepaniak}
\author[Katow]{E.~Stephan}
\author[IKPJ]{G.~Sterzenbach}
\author[IKPJ]{H.~Stockhorst}
\author[IKPJ,JARA]{H.~Str\"oher}
\author[IFJ]{A.~Szczurek}
\author[ASWarsN]{A.~Trzci{\'n}ski}
\author[IITB]{R.~Varma}
\author[IKPUU]{M.~Wolke}
\author[IPJ]{A.~Wro{\'n}ska}
\author[ZELJ]{P.~W\"ustner}
\author[KEK]{A.~Yamamoto}
\author[ASLodz]{J.~Zabierowski}
\author[IPJ]{M.J.~Zieli{\'n}ski}
\author[IKPUU]{J.~Z{\l}oma{\'n}czuk}
\author[ASWarsN]{P.~{\.Z}upra{\'n}ski}
\author[IKPJ]{M.~{\.Z}urek}
\author[HISKP]{and B.~Kubis}
\author[IKPUU]{S.~Leupold}

\address[IKPUU]{Division of Nuclear Physics, Department of Physics and 
 Astronomy, Uppsala University, Box 516, 75120 Uppsala, Sweden}
\address[ASWarsN]{Department of Nuclear Physics, National Centre for Nuclear 
 Research, ul.\ Hoza~69, 00-681, Warsaw, Poland}
\address[IPJ]{Institute of Physics, Jagiellonian University, prof.\ 
 Stanis{\l}awa {\L}ojasiewicza~11, 30-348 Krak\'{o}w, Poland}
\address[Edinb]{School of Physics and Astronomy, University of Edinburgh, 
 James Clerk Maxwell Building, Peter Guthrie Tait Road, Edinburgh EH9 3FD, 
 Great Britain}
\address[MS]{Institut f\"ur Kernphysik, Westf\"alische Wilhelms--Universit\"at 
 M\"unster, Wilhelm--Klemm--Str.~9, 48149 M\"unster, Germany}
\address[ASWarsH]{High Energy Physics Department, National Centre for Nuclear 
 Research, ul.\ Hoza~69, 00-681, Warsaw, Poland}
\address[IITB]{Department of Physics, Indian Institute of Technology Bombay, 
 Powai, Mumbai--400076, Maharashtra, India}
\address[Budker]{Budker Institute of Nuclear Physics of SB RAS, 11~akademika 
 Lavrentieva prospect, Novosibirsk, 630090, Russia}
\address[Novosib]{Novosibirsk State University, 2~Pirogova Str., Novosibirsk, 
 630090, Russia}
\address[IKPJ]{Institut f\"ur Kernphysik, Forschungszentrum J\"ulich, 52425 
 J\"ulich, Germany}
\address[IFJ]{The Henryk Niewodnicza{\'n}ski Institute of Nuclear Physics, 
 Polish Academy of Sciences, 152~Radzikowskiego St, 31-342 Krak\'{o}w, Poland}
\address[PITue]{Physikalisches Institut, Eberhard--Karls--Universit\"at 
 T\"ubingen, Auf der Morgenstelle~14, 72076 T\"ubingen, Germany}
\address[Kepler]{Kepler Center for Astro and Particle Physics, Eberhard Karls 
 University T\"ubingen, Auf der Morgenstelle~14, 72076 T\"ubingen, Germany}
\address[ZELJ]{Zentralinstitut f\"ur Engineering, Elektronik und Analytik, 
 Forschungszentrum J\"ulich, 52425 J\"ulich, Germany}
\address[Erl]{Physikalisches Institut, Friedrich--Alexander--Universit\"at 
 Erlangen--N\"urnberg, Erwin--Rommel-Str.~1, 91058 Erlangen, Germany}
\address[ITEP]{Institute for Theoretical and Experimental Physics, State 
 Scientific Center of the Russian Federation, Bolshaya Cheremushkinskaya~25, 
 117218 Moscow, Russia}
\address[Giess]{II.\ Physikalisches Institut, Justus--Liebig--Universit\"at 
 Gie{\ss}en, Heinrich--Buff--Ring~16, 35392 Giessen, Germany}
\address[IITI]{Department of Physics, Indian Institute of Technology Indore, 
 Khandwa Road, Indore--452017, Madhya Pradesh, India}
\address[HepGat]{High Energy Physics Division, Petersburg Nuclear Physics 
 Institute, Orlova Rosha~2, Gatchina, Leningrad district 188300, Russia}
\address[HiJINR]{Veksler and Baldin Laboratory of High Energiy Physics, Joint 
 Institute for Nuclear Physics, Joliot--Curie~6, 141980 Dubna, Moscow region, 
 Russia}
\address[Katow]{August Che{\l}kowski Institute of Physics, University of 
 Silesia, Uniwersytecka~4, 40-007, Katowice, Poland}
\address[NITJ]{Department of Physics, Malaviya National Institute of 
 Technology Jaipur, 302017, Rajasthan, India}
\address[JARA]{JARA--FAME, J\"ulich Aachen Research Alliance, Forschungszentrum 
 J\"ulich, 52425 J\"ulich, and RWTH Aachen, 52056 Aachen, Germany}
\address[Bochum]{Institut f\"ur Experimentalphysik I, Ruhr--Universit\"at 
 Bochum, Universit\"atsstr.~150, 44780 Bochum, Germany}
\address[Tomsk]{Department of Physics, Tomsk State University, 36~Lenina 
 Avenue, Tomsk, 634050, Russia}
\address[KEK]{High Energy Accelerator Research Organisation KEK, Tsukuba, 
 Ibaraki 305--0801, Japan}
\address[ASLodz]{Department of Astrophysics, National Centre for Nuclear 
 Research, Box~447, 90--950 {\L}\'{o}d\'{z}, Poland}
\address[HISKP]{Helmholtz--Institut f\"ur Strahlen-- und Kernphysik, 
 Rheinische Friedrich--Wilhelms--Universit\"at Bonn, Nu{\ss}allee~14--16, 
 53115 Bonn, Germany}

\fntext[fnmz]{present address: Institut f\"ur Kernphysik, Johannes 
 Gutenberg--Universit\"at Mainz, Johann--Joachim--Becher Weg~45, 55128 Mainz, 
 Germany}
\fntext[fnpgi]{present address: Peter Gr\"unberg Institut, PGI--6 Elektronische 
 Eigenschaften, Forschungszentrum J\"ulich, 52425 J\"ulich, Germany}
\fntext[fndus]{present address: Institut f\"ur Laser-- und Plasmaphysik, 
 Heinrich--Heine Universit\"at D\"usseldorf, Universit\"atsstr.~1, 40225 
 Düsseldorf, Germany}
\fntext[fnac]{present address: III.~Physikalisches Institut~B, Physikzentrum, 
 RWTH Aachen, 52056 Aachen, Germany}
\fntext[fnjcns]{present address: J\"ulich Centre for Neutron Science JCNS, 
 Forschungszentrum J\"ulich, 52425 J\"ulich, Germany}
\fntext[fnharv]{present address: Department of Physics, Harvard University, 
 17~Oxford St., Cambridge, MA~02138, USA}
\fntext[fnlnf]{present address: INFN, Laboratori Nazionali di Frascati, Via 
 E.~Fermi, 40, 00044 Frascati (Roma), Italy}

\cortext[coau]{Corresponding authors }

\begin{abstract}

Using the production reactions $pd\to {}^3\mbox{He}\,\omega$ and $pp\to pp\omega$, the Dalitz plot distribution for the $\omega \to \pi^+ \pi^- \pi^0$ decay is studied with the WASA detector at COSY, based on a combined data sample of $ (4.408\pm 0.042) \times 10^4$ events.  The Dalitz plot density is parametrised by a product of the $P$-wave phase space and a polynomial expansion in the normalised polar Dalitz plot variables $Z$ and $\phi$. For the first time, a deviation from pure $P$-wave phase space is observed with a significance of $4.1\sigma$. The deviation is parametrised by a linear term $1+2\alpha Z$, with $\alpha$ determined to be $+0.147\pm0.036$, consistent with the
expectations of $\rho$-meson-type final-state interactions of the $P$-wave pion pairs. 

\end{abstract}

\begin{keyword}
Decays of other mesons\sep Meson--meson interactions\sep Light mesons

\PACS 13.25.Jx\sep 
13.75.Lb\sep 14.40.Be
\end{keyword}

\end{frontmatter}

\section{Introduction}

The present work and foreseeable follow-ups are based on two motivations: 1.\ To check and improve on our understanding of the importance of hadronic final-state interactions for the structure and decays of hadrons. 2.\ To improve the Standard Model prediction for the gyromagnetic ratio of the muon~\cite{Colangelo:2014pva}. The present work accomplishes the first task concerning the Dalitz decay of the $\omega$ meson into three pions; it also constitutes a significant step forward towards providing improved hadronic input for the second task. In the following we shall discuss the two tasks in more detail.

The $\omega$-meson resonance was discovered in 1961~\cite{Maglic:1961nz}. Its main decay branch is $\omega \to \pi^+ \pi^- \pi^0$, with a branching ratio of $\text{BR} = (89.2\pm0.7)\%$. By now it is well established that the $\omega$ meson has spin-parity $J^P = 1^-$~\cite{Agashe:2014kda}. As a consequence, the combination of Bose, isospin, and parity symmetry of the strong interaction demands that for the decay $\omega \to \pi^+ \pi^- \pi^0$ every pion pair is in a state of odd relative orbital angular momentum. Given the limited phase space of the decay one can safely assume the $P$-wave to be the dominant partial wave.\footnote{Genuine $F$-wave corrections have been  modelled theoretically, and found to be tiny~\cite{Niecknig:2012sj}.} If a pion pair is in a $P$-wave state, then the third pion will be in $P$-wave state relative to the pair. This ``$P$-wave phase space'' distribution has been confirmed experimentally. Historically the $P$-wave dominance of the decay has actually been used to pin down the quantum numbers of the $\omega$ meson~\cite{Stevenson:1962zz,Alff:1962zz,Xuong:1962zz,Danburg:1971ui}. 

If the pions, once produced in the decay, did not interact further, then solely the $P$-wave phase space would shape the Dalitz plot of the decay $\omega \to \pi^+ \pi^- \pi^0$. However, a pion pair in a $P$-wave shows a very strong final-state interaction. The two-pion $P$-wave phase shift is dominated by the $\rho$ meson, and is now known very accurately~\cite{Ananthanarayan:2000ht,Colangelo:2001df,GarciaMartin:2011cn}; this is essential in particular for theoretical studies of these decays using dispersion theory, which use the phase shifts as input directly~\cite{Niecknig:2012sj,Danilkin:2014cra}. In the similar decay $\phi \to \pi^+ \pi^- \pi^0$ one can see the $\rho$ meson as a resonance in the corresponding Dalitz plot~\cite{Aloisio:2003ur,Akhmetshin:2006sc}. In the decay $\omega \to \pi^+ \pi^- \pi^0$ there is not enough energy for a pion pair to reach the $\rho$ resonance mass; yet already  for the available invariant masses the two-pion $P$-wave phase shift is significantly different from zero. In fact, every theoretical approach that deals with the decay $\omega \to \pi^+ \pi^- \pi^0$ includes this non-trivial phase shift and/or the $\rho$ meson in one way or the other; see, {\it e.g.}, Refs.~\cite{GellMann:1962jt,Klingl:1996by,Harada:2003jx,RuizFemenia:2003hm,Leupold:2008bp,Niecknig:2012sj,Terschlusen:2013iqa,Danilkin:2014cra} 
and references therein. In practice this leads to an increase of strength towards the boundaries of the Dalitz plot, superimposed with the pure $P$-wave phase space, which drops towards the boundaries. This increase of strength is on the level of about 20\%~\cite{Leupold:2008bp,Niecknig:2012sj}. 

The prediction of the presence of final-state interactions leading to an increase of strength towards the Dalitz-plot boundaries ought to be tested experimentally. Interestingly this has not been achieved so far. The highest statistics of a dedicated $\omega \to \pi^+ \pi^- \pi^0$ Dalitz plot measurement from 1966 had $4208\pm75$ signal events~\cite{Flatte:1966zza}. Due to the limited statistics, fits with a pure $P$-wave phase space could not be distinguished from a distortion by the final-state interactions, {\it i.e.}\ by intermediate $\rho\pi$ states. Surprisingly there were no further dedicated Dalitz plot studies of the $\omega \to \pi^+ \pi^- \pi^0$ decay. In the present work we will reveal that the universal final-state interactions of the pion pairs are indeed present in the $\omega$ Dalitz decay. In the analysis presented here we have produced an acceptance-corrected Dalitz plot and extracted experimental values for parameters describing the density distribution. This constitutes the first task spelled out in the beginning of this introduction.

The second motivation for a precise measurement of the $\omega\to \pi^+ \pi^- \pi^0$ Dalitz plot consists in improving hadronic input for the theoretical assessment of the hadronic light-by-light scattering contribution to the anomalous magnetic moment of the muon.  The largest individual contribution is given by the lightest hadronic intermediate state, the so-called $\pi^0$ pole term, whose strength is determined by the corresponding singly and doubly virtual transition form factors. One of the few possibilities to gain experimental access to the \textit{doubly} virtual $\pi^0$ transition form factor with high precision consists in studying vector meson conversion decays, in particular $\omega\to\pi^0\ell^+\ell^-$---which is intimately linked, through dispersion relations, to the $\omega\to \pi^+ \pi^- \pi^0$ decay amplitude~\cite{Koepp:1974da,Schneider:2012ez,Danilkin:2014cra}. However, these theoretical descriptions of the $\omega$ transition form factor (see also Ref.~\cite{Terschluesen:2010ik})  fail to describe the very precise data on $\omega\to\pi^0\mu^+\mu^-$ taken by the NA60 collaboration~\cite{Arnaldi:2009aa,Arnaldi:2016pzu}, which may violate very fundamental theoretical bounds~\cite{Ananthanarayan:2014pta,Caprini:2015wja}. In all these studies, the $\omega\to \pi^+ \pi^- \pi^0$ decay amplitude is a potential loose end, as it is so far only theoretically modelled, not experimentally tested. In combination with the precisely measured $\phi\to \pi^+ \pi^- \pi^0$ Dalitz plot information, it could be used to further constrain the amplitude analysis of $e^+e^-\to \pi^+ \pi^- \pi^0$, and hence the $\pi^0$ transition form factor in a wider range~\cite{Hoferichter:2014vra}.

Recently, an easy-to-use polynomial parametrisation of the $\omega \to \pi^+ \pi^- \pi^0$ Dalitz plot distribution has been suggested~\cite{Niecknig:2012sj}, as a generalisation of the commonly used one for the decay $\eta \to 3\pi^0$ (which has a similar crossing symmetry). In the present work we utilise the same parametrisation and compare to recent theoretical approaches~\cite{Niecknig:2012sj,Terschlusen:2013iqa,Danilkin:2014cra} that  have provided predictions for the corresponding Dalitz plot parameters. Refs.~\cite{Niecknig:2012sj,Danilkin:2014cra} are based on dispersion theory: both employ the pion--pion $P$-wave scattering  phase shift as input and describe rescattering of all three final-state pions consistently to all orders. The Dalitz plot parameters cited from Ref.~\cite{Danilkin:2014cra} in Table~\ref{tab:DPthfitpar} refer to a formalism with only the overall normalisation of the amplitude fixed from experiment, hence the energy dependence is fully predicted; the ranges as given in Ref.~\cite{Niecknig:2012sj} include (in addition to variations in the phase shift input etc.) potential deviations from such  a pure prediction as gleaned from the analysis of the analogous $\phi\to\pi^+\pi^-\pi^0$ Dalitz plot~\cite{Aloisio:2003ur}. Ref.~\cite{Terschlusen:2013iqa} is based on an effective Lagrangian for the lightest pseudoscalar and vector mesons. The strength of the initial $\omega$-$\rho$-$\pi$ interaction is fitted to the decay width of $\omega \to \pi^+\pi^-\pi^0$ and cross-checked with the decay width of $\omega \to \pi^0\gamma$. The Lagrangian provides the kernel for a Bethe--Salpeter equation that generates the two-pion rescattering. In contrast to the dispersive approaches, crossed-channel rescattering of the three-pion system is not included.

\begin{table}
\centering
\renewcommand{\arraystretch}{1.2}
\begin{tabular}{lrr}
\toprule
 & $\alpha\times 10^3$& $\beta\times 10^3$\\
\midrule
Uppsala~\cite{Terschlusen:2013iqa} &202& --\\
Bonn~\cite{Niecknig:2012sj} &$84\ldots102$& --\\
JPAC~\cite{Danilkin:2014cra} & 94& -- \\
\midrule
Uppsala& 190& 54\\
Bonn &$74\ldots90$& $24\ldots30$\\
JPAC &84& 28\\
\bottomrule
\end{tabular}
\caption{The Dalitz plot parameters from fits to the three theoretical predictions~\cite{Terschlusen:2013iqa,Niecknig:2012sj,Danilkin:2014cra}, where at most two parameters were used in the fit.  \label{tab:DPthfitpar}}
\renewcommand{\arraystretch}{1.0}
\end{table}

One way to describe a three-particle Dalitz decay distribution is to use invariant masses of particle pairs~\cite{Agashe:2014kda}. This would be particularly useful for reading off resonance masses if the decay was mediated by one or several resonances. However, there are no intermediate resonances in the kinematically accessible energy range of the decay $\omega \to \pi^+ \pi^- \pi^0$ that would be compatible with the symmetries of the strong interaction. For our case of interest, we first split off the $P$-wave phase space (see, {\it e.g.}, Refs.~\cite{Leupold:2008bp,Niecknig:2012sj}) and parametrise the rest by a polynomial distribution following~\cite{Niecknig:2012sj}. Denoting the polarisation vector of the $\omega$ meson by $\epsilon(P_\omega,\lambda_\omega)$ and the momenta of the outgoing pions by $P_+$, $P_-$, and $P_0$, we start with the most general matrix element compatible with the symmetries,

\begin{equation}
{\cal M} = i \varepsilon_{\mu\nu\alpha\beta} \, \epsilon^\mu \, P_+^\nu \, P_-^\alpha \, P_0^\beta \, {\cal F}  \,.
\label{eq:matrelem}  
\end{equation}

The dynamics of the final-state interactions is encoded in the scalar function ${\cal F}$~\cite{Leupold:2008bp,Niecknig:2012sj}.
After summation over the helicity $\lambda_\omega$ of the $\omega$ meson one obtains a Dalitz plot distribution proportional to

\begin{equation}
\sum\limits_{\lambda_\omega} \vert {\cal M}\vert^2 \propto {\cal P} \, \vert {\cal F}\vert^2
\label{eq:distr-w-p}  
\end{equation}

with the pure $P$-wave phase-space distribution

\begin{align}
{\cal P} & = m_+^2m_-^2m_0^2+2(P_+P_-)(P_-P_0)(P_0P_+)
\nonumber \\  &
- m_+^2(P_-P_0)^2-m_-^2(P_+P_0)^2-m_0^2(P_+P_-)^2   \,.
\label{eq:pwave}
\end{align}

Note that for the ${\cal P}$ term we can account for ``kinematic'' isospin violations due to the difference between the masses of the uncharged and charged pions, $m_0$ and $m_\pm$, respectively. For the remaining distribution ${\cal F}$, which covers the dynamics of the final-state interaction, we ignore isospin breaking effects.

The quantity ${\cal F}$ and therefore also $\vert {\cal F}\vert^2$ would be a constant if there were no final-state interactions between the produced pions. In reality $\vert {\cal F}\vert^2$ is not a constant, but relatively flat. Instead of parametrising $\vert {\cal F}\vert^2$  by invariant masses of pion pairs we follow Ref.~\cite{Niecknig:2012sj} and utilise normalised variables $X$ and $Y$, which have their origin at the centre of the Dalitz plot. They are defined by

 \begin{equation}
X = \sqrt{3}\ \frac{T_{+} - T_{-}}{Q_\omega} ,\qquad
Y = \frac{3T_{0}}{Q_\omega} -1 \label{eq:YasT} ,
\end{equation}

with

\begin{equation}
Q_\omega =T_{+} +  T_{-} + T_{0} . \label{eq:Q}
\end{equation}
Here $T_i$ are the kinetic energies of the pions in the $\omega$ rest frame (centre-of-mass frame of the three-pion system). Finally one introduces polar coordinates by

\begin{equation}\label{eq:ZPhiDef}
X = \sqrt{Z}\cos\phi,\quad Y = \sqrt{Z}\sin\phi \,.
\end{equation}

The expansion for $\vert {\cal F}\vert^2$, valid in the isospin limit, reads

\begin{equation}
\vert {\cal F}\vert^2(Z,\phi) = {\cal N}\cdot{\cal G}(Z,\phi) \,,
\end{equation}

where ${\cal N}$ is a normalisation constant and ${\cal G}$ contains the expansion in $Z$ and $\phi$~\cite{Niecknig:2012sj}:

\begin{equation}\label{eq:DPamplitude}
{\cal G}(Z,\phi) = 1+ 2\alpha Z+ 2\beta Z^{3/2}\sin3\phi+ 2\gamma Z^2+\mathcal{O}\big(Z^{5/2}\big)  \, .
\end{equation}

The Dalitz plot distribution can then be fitted using this formula to
extract the ``Dalitz plot parameters'' \mbox{$\alpha$, $\beta$, $\gamma$, \ldots}.
The fit results to the theory predictions of Refs.~\cite{Niecknig:2012sj,Terschlusen:2013iqa,Danilkin:2014cra}, if Eq.~\eqref{eq:DPamplitude} is truncated at order $Z$ (one parameter fit) or at order $Z^{3/2}$ (two parameter fit), are shown in Table~\ref{tab:DPthfitpar}.  The reproduction of the theoretical Dalitz plot distributions is improved significantly in all cases when including the term $\propto \beta$.

\section{The experiment}

The experimental data was collected using the WASA setup, where the $\omega$ was produced in the $pd\to{}^3\textrm{He}\,\omega$ reaction and in the $pp\to pp\omega$ reaction. The WASA detector~\cite{Bargholtz:2008aa,Adam:2004ch} is an internal target experiment at the Cooler Synchrotron (COSY) storage ring, Forschungszentrum J\"ulich, Germany.  The COSY proton beam interacts with an internal target consisting of small pellets of frozen hydrogen or deuterium (diameter $\sim 35\,\mu$m).

The WASA detector consists of a Central Detector (CD) and a Forward Detector (FD), covering scattering angles of 20$^{\circ}$--169$^{\circ}$ and 3$^{\circ}$--18$^{\circ}$, respectively.  The CD is used to measure decay products of the mesons.  A cylindrical straw chamber (MDC) is placed in a magnetic field of 1\,T, provided by a superconducting solenoid. The electromagnetic calorimeter (SEC) consists of 1012 CsI(Na) crystals which are read out by photomultipliers. A plastic scintillator barrel (PSB) is placed between the MDC and the SEC, allowing particle identification and accurate timing for charged particles.  The FD consists of thirteen layers of plastic scintillators for energy and time
determination and a straw tube tracker providing a precise track direction.

\begin{figure}
\centering 
\includegraphics[width=0.7\linewidth]{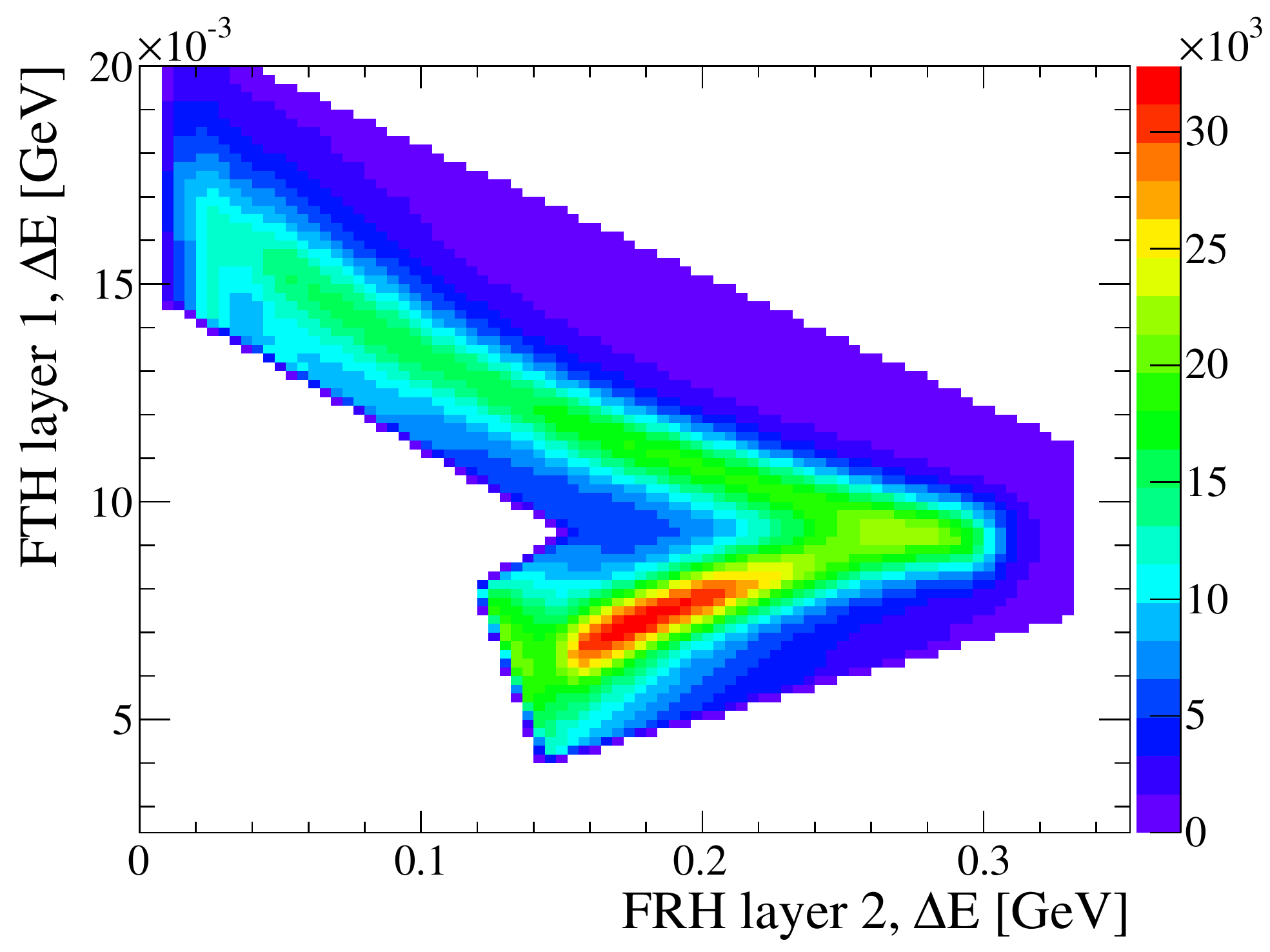}
\includegraphics[width=0.7\linewidth]{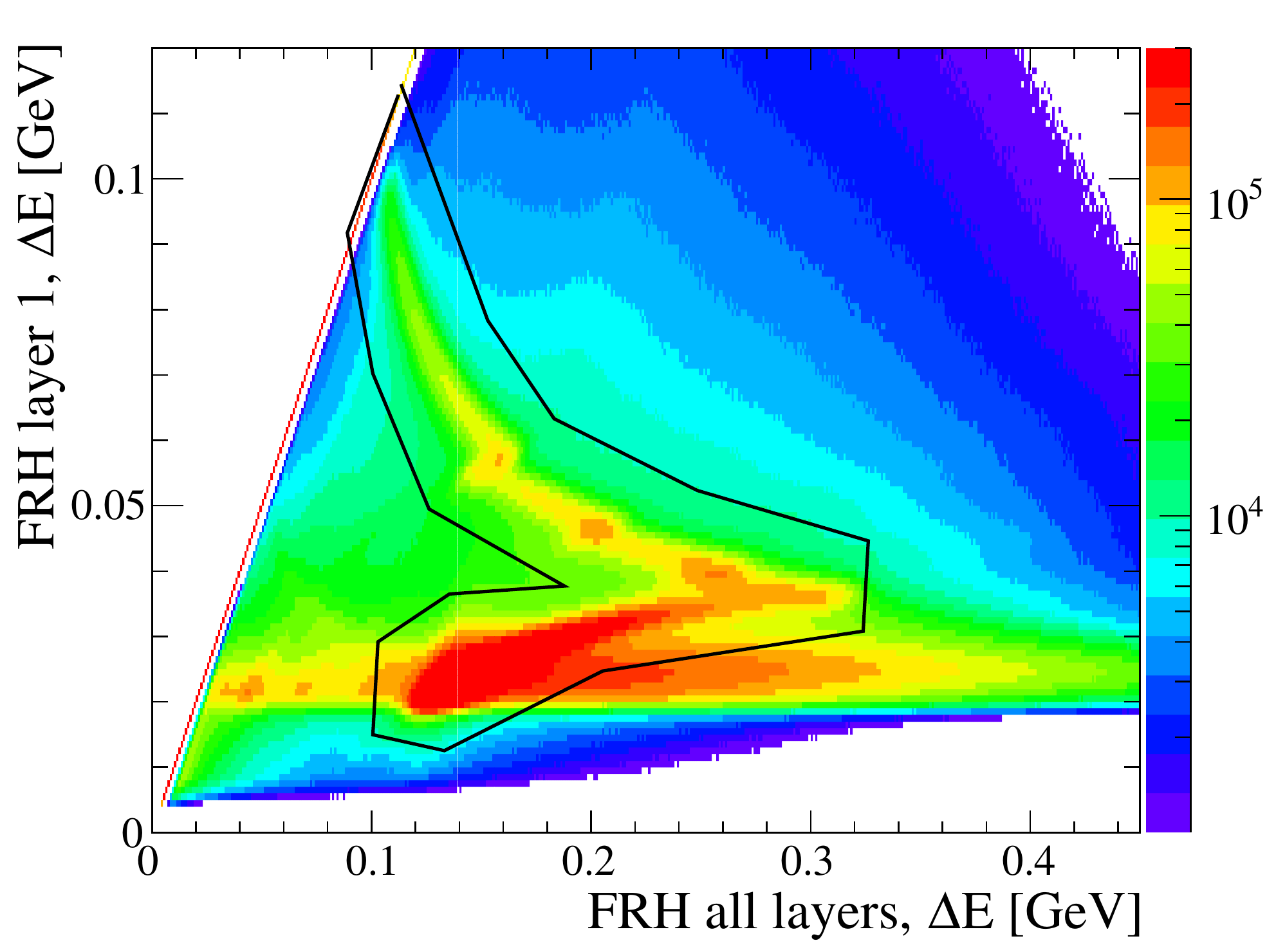}
\caption{(top) Correlation of energy  deposits in two Forward Detector  plastic detector layers for ${}^3$He identification: the first thick layer (11  cm) and a preceding thin (0.5 cm) layer. (bottom)  Correlation of energy  deposits in several Forward Detector plastic detector layers for proton identification. The black line shows the selection region for proton candidates.} \label{FVC} 
\end{figure}

When the $\omega$ mesons were produced using the $pd\to\mbox{}^{3}\mbox{He}\,\omega$ reaction, two different proton kinetic energies were used: $T_A=1.450\,\mathrm{GeV}$ and $T_B=1.500\,\mathrm{GeV}$. The cross section of the reaction is 84(10)~nb at the lower energy~\cite{Schonning:2009nf} and was studied previously by the CELSIUS/WASA collaboration. Triggers select events  with at least  one track in  the FD
with a high energy deposit in the thin plastic scintillator layers. This condition allows for an efficient selection of $^3$He  ions and provides an  unbiased data sample  of $\omega$ meson decays. The proton beam energy was chosen so that the $\mbox{}^{3}\mbox{He}$ produced in the $pd\rightarrow\mbox{}^{3}\mbox{He}\,\omega$ reaction stops in the second thick scintillator layer of the FD. The correlation  plot $\Delta  E- \Delta  E$ from a  thin layer and  the first thick layer  of the  FD is shown  in Fig.~\ref{FVC}(top). The  band corresponding to the $\mbox{}^{3}\mbox{He}$ ion is well separated from the bands for  other particles and allows a  clear identification of $\mbox{}^{3}\mbox{He}$. The $\mbox{}^{3}\mbox{He}$ from the reaction of interest has kinetic energies up to 700\,MeV and scattering angles ranging from $0^{\circ}$ to $10^{\circ}$.

The $pp\to pp\omega$ experiment was performed at $T_C=2.063$\,GeV beam kinetic energy, corresponding to 60\,MeV centre-of-mass excess energy and cross section 5.7\,$\mu$b~\cite{Barsov:2006sc}. In the $pp$ collision experiment, the selected events were required, at trigger level, to contain at least two tracks reaching the second thick layer of the plastic scintillators in the FD, at least two hits in the PSB, and at least one cluster in the SEC. In the offline analysis, pairs of tracks corresponding to the $\Delta E-\Delta E$ proton bands, shown in Fig.~\ref{FVC}(bottom), in different thick layers of the FD are selected as proton pair candidates.

For the particles measured in the CD, a common analysis procedure is used for all three data sets. Events are selected if they contain at least one pair of opposite charge particle tracks in the MDC with scattering angles greater than 30$^\circ$ and at least two neutral clusters with energy deposit above 20\,MeV in the SEC.  Relative time between the tracks is checked to minimise pile ups. The charged particle tracks are assigned the charged pion mass. Possible combinations of the charged and neutral particle tracks for the selected events are tested using a constrained kinematic fit assuming the conservation of energy and momentum with the $pd\to{}^3\mbox{He}\,\pi^+\pi^-\gamma\gamma$ or $pp\to pp\pi^+\pi^-\gamma\gamma$ hypothesis, respectively. The events with $p$-values less than 0.05 are rejected. For the case when more than one track combination fulfils the selection, the combination giving larger $p$-value is selected. Finally, further background suppression is achieved by applying a kinematic fit with the contending hypothesis $pd\to{}^3\mbox{He}\,\pi^+\pi^-$ or $pp\to pp\pi^+\pi^-$, respectively. If the resulting $p$-value is larger than for the first fit, the event is rejected.

\begin{figure*}
\centering
\subfigure[]{\includegraphics*[width=0.32\linewidth]{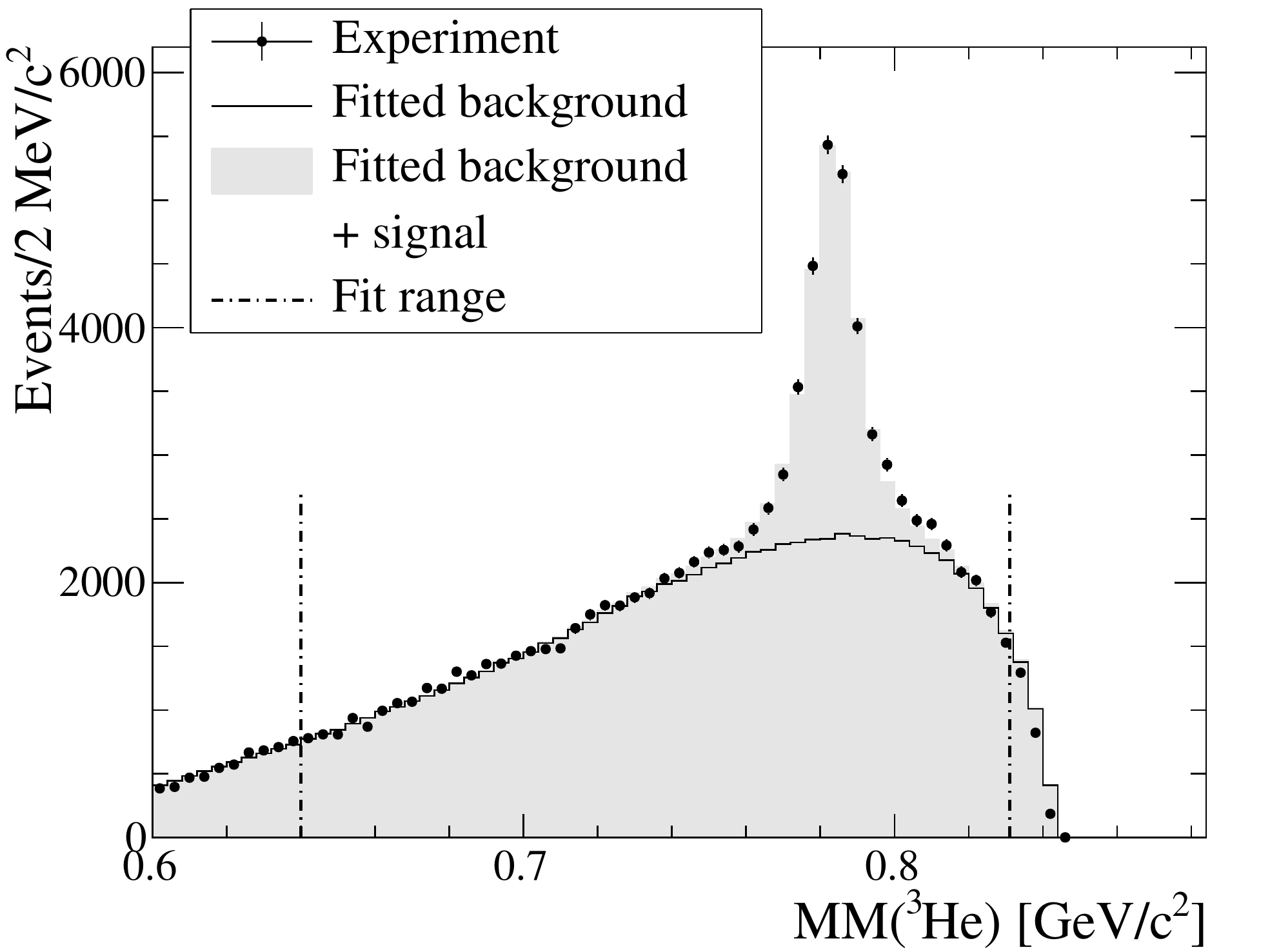}} \hfill
\subfigure[]{\includegraphics*[width=0.32\linewidth]{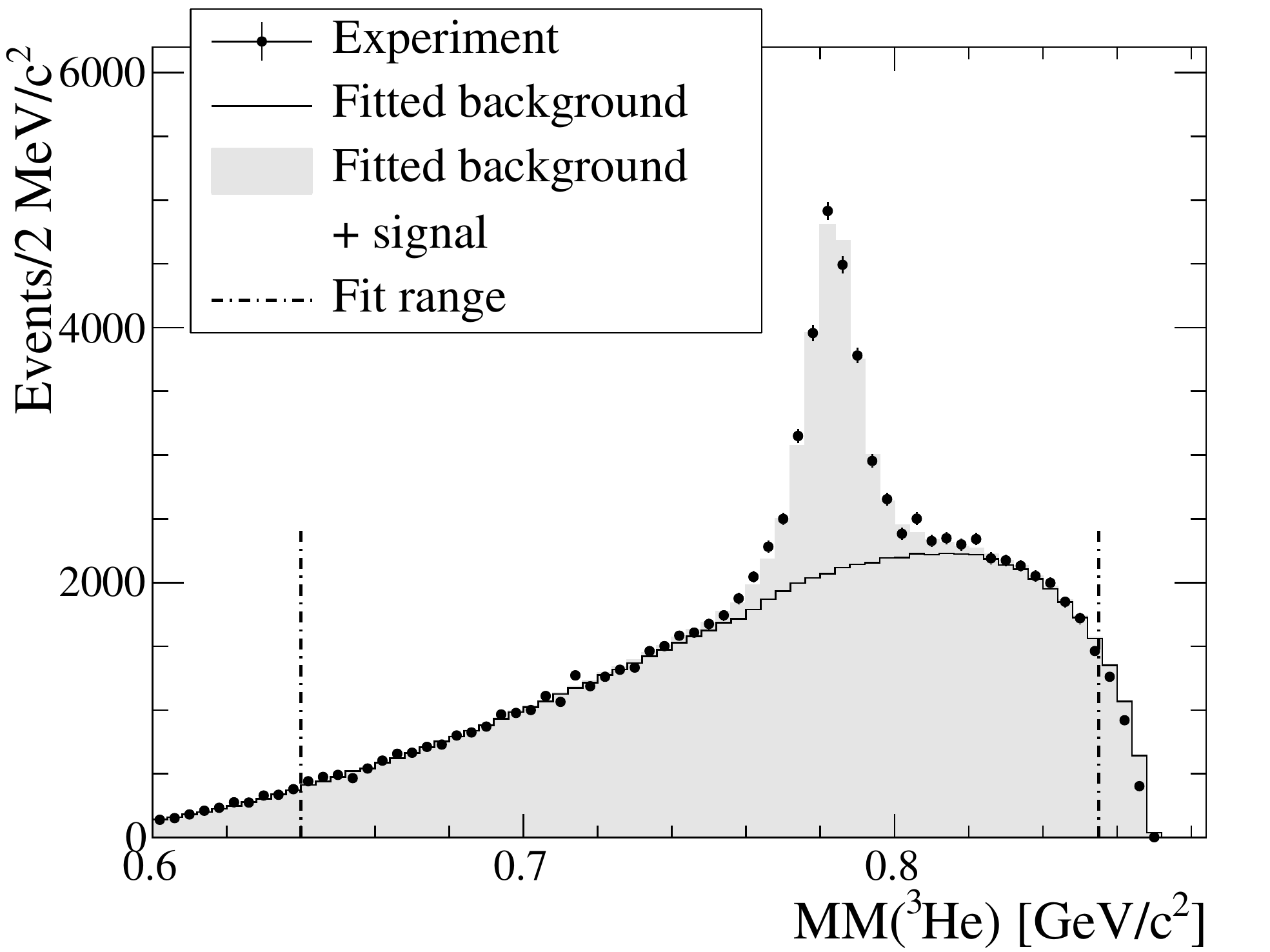}} \hfill
\subfigure[]{\includegraphics*[width=0.32\linewidth]{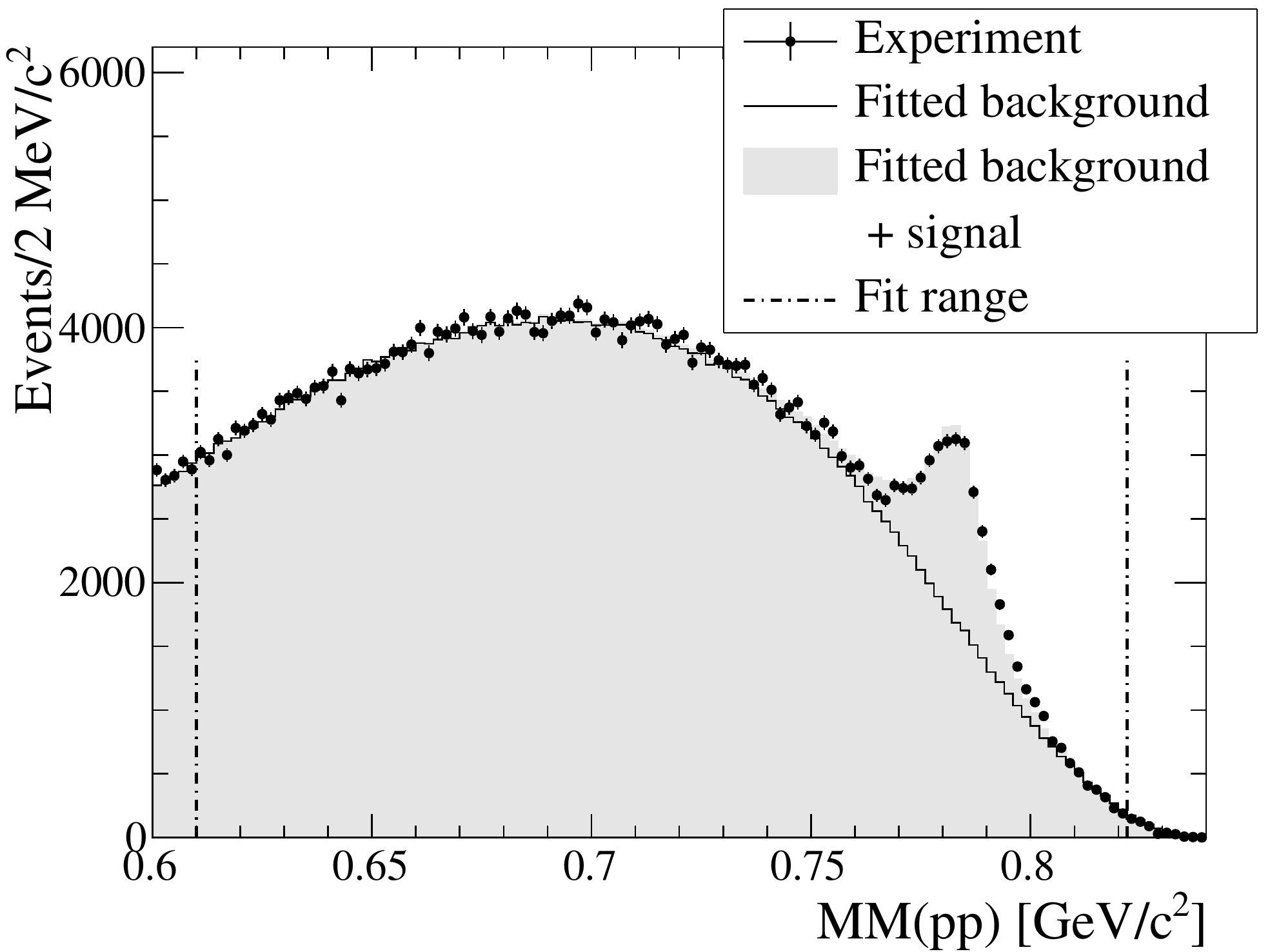}}
\caption{Missing mass distributions after the full analysis procedure as well as the result of the fit Eq.~\eqref{eq:MMFitFunc}.  (a): T${}_{A}=1.450\, \textrm{GeV}$. (b): T${}_{B}=1.500\, \textrm{GeV}$. (c): T${}_{C}=2.063\, \textrm{GeV}$.}\label{fig:MMFit}
\end{figure*}

The missing mass distributions, MM$({}^3\textrm{He})$ and MM$(pp)$, for the three data sets are shown in Fig.~\ref{fig:MMFit}.  The missing masses, calculated from the variables corrected by the kinematic fit, are equivalent to the invariant mass of the $\pi^+\pi^-\gamma\gamma$ system.
The observed $\omega$ peak position is shifted from the nominal value of the $\omega$ mass by $+$0.7\,MeV for MM$({}^3\textrm{He})$ in the two $pd$ data sets and by $+$1.1\,MeV for MM$(pp)$. The observed shifts correspond to deviations from the nominal beam energy by 0.55\,MeV and 0.75\,MeV, respectively, which is well within the uncertainty of the absolute energy scale of COSY. To reproduce the experimental $\omega$ peak position, the missing mass distributions were shifted accordingly. To also reach agreement between experiment and simulation for the width of the $\omega$ peak, the resolution from simulated detector responses were adjusted.

Both the background shape and the $\omega$ peak content are fitted simultaneously to the experimental distribution using the following fit function:

\begin{equation}
H(\mu) = N_S H_{\omega}(\mu)+\big\{a_0 + a_1\mu + a_2\mu^2+ a_3\mu^3 \big\}\times H_{3\pi}(\mu) , \label{eq:MMFitFunc}
\end{equation}

where $\mu = \textrm{MM}({}^3\textrm{He})$ or $\textrm{MM}(pp)$. $H_{\omega}(\mu)$ and $H_{3\pi}(\mu)$ represent reconstructed distributions of simulated signal and background and correspond to events that have passed through the same analysis steps as the experimental data. $H_{\omega}(\mu)$ is normalised such that the fit gives directly the number of signal events, $N_S$, and the related error. The other parameters fitted are $a_0$, $a_1$, $a_2$, and $a_3$ (in case of $pd$ data $a_3$ is set to 0). The range in $\mu$ used for the fit is $[0.640,0.832]$\,GeV/c${}^2$ for set~A, $[0.640,0.856]$\,GeV/c${}^2$ for set~B, and  $[0.608, 0.824]$\,GeV/c${}^2$ for set~C. The limits of these ranges are shown by the dashed lines in Fig.~\ref{fig:MMFit}, where the result of these fits to the full data samples are given. The resulting number of events is: 14600(200) for set~A, 13500(200) for set~B, and 16000(300) for set~C.

\section{Dalitz plot}

\begin{figure}
\centering
\includegraphics[width=0.9\linewidth]{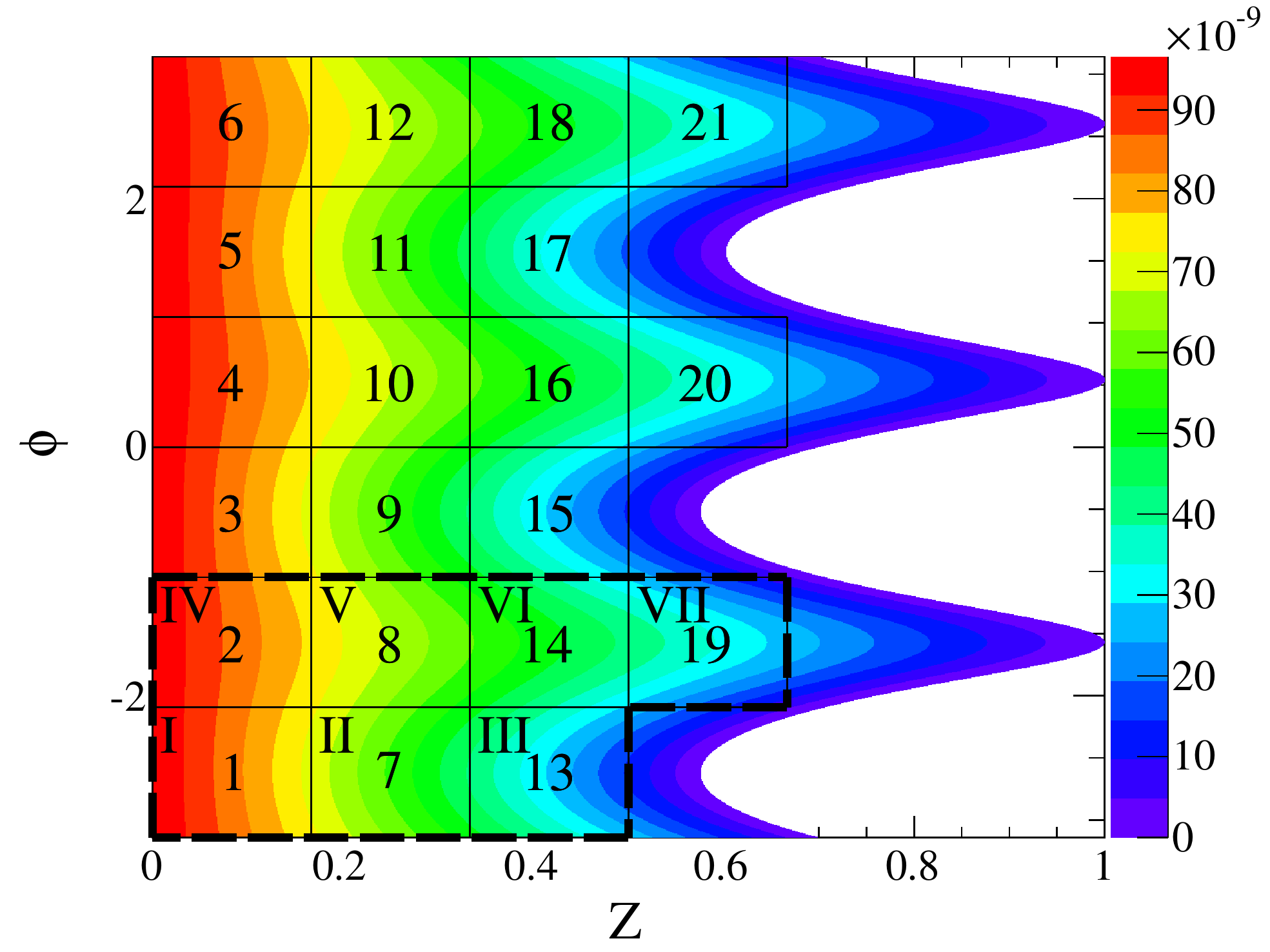}
\caption{Arabic numbering of the bins used in the analysis. Roman numbering of the sectors used in consistency checks. The colour plot shows the kinematically allowed region of the $\omega\to\pi^+\pi^-\pi^0$ reaction with $\omega$ nominal mass as well as the density distribution from $P$-wave dynamics.}\label{fig:DPbins}
\end{figure}

The Dalitz plot density is represented using a two-dimensional histogram in the $Z$ and $\phi$ variables, defined in Eq.~\eqref{eq:ZPhiDef}. The size of the selected bins is determined by the experimental resolution of $Z$ and $\phi$ and the statistics of the collected data sample. The number of events in each bin should be sufficient for determining the signal yield and to carry out a $\chi^2$ fit of the Dalitz plot density parametrisation.  The $\phi$ variable range $[-\pi,\pi]$ is divided into six bins to preserve the threefold isospin symmetry and to be sensitive to a possible $\sin 3\phi$ dependence. The $Z$ variable range $[0,1]$ is also divided into six bins.  Only the 21 bins fully contained inside the kinematic limits of the decay are used. Figure~\ref{fig:DPbins} introduces the bin numbering used for the presentation of the results. 

A small shift of the Dalitz plot along the $Y$ axis is due to the mass difference between the neutral and charged pions. It is most visible in Fig.~\ref{fig:DPbins} when comparing the regions at $\phi=\pi/2$ to the ones at $\phi=-\pi/6$ and $-2\pi/3$. The picture shows also seven sectors I--VII that are used to test the consistency of the fit results.

For each Dalitz plot bin, the experimental missing mass distribution is constructed and the number of entries in the $\omega$ peak is extracted by fitting a simulated $\omega\to\pi^+\pi^-\pi^0$ signal along with background contributions using Eq.~\eqref{eq:MMFitFunc}.

\begin{figure}
\centering
\includegraphics*[width=\linewidth]{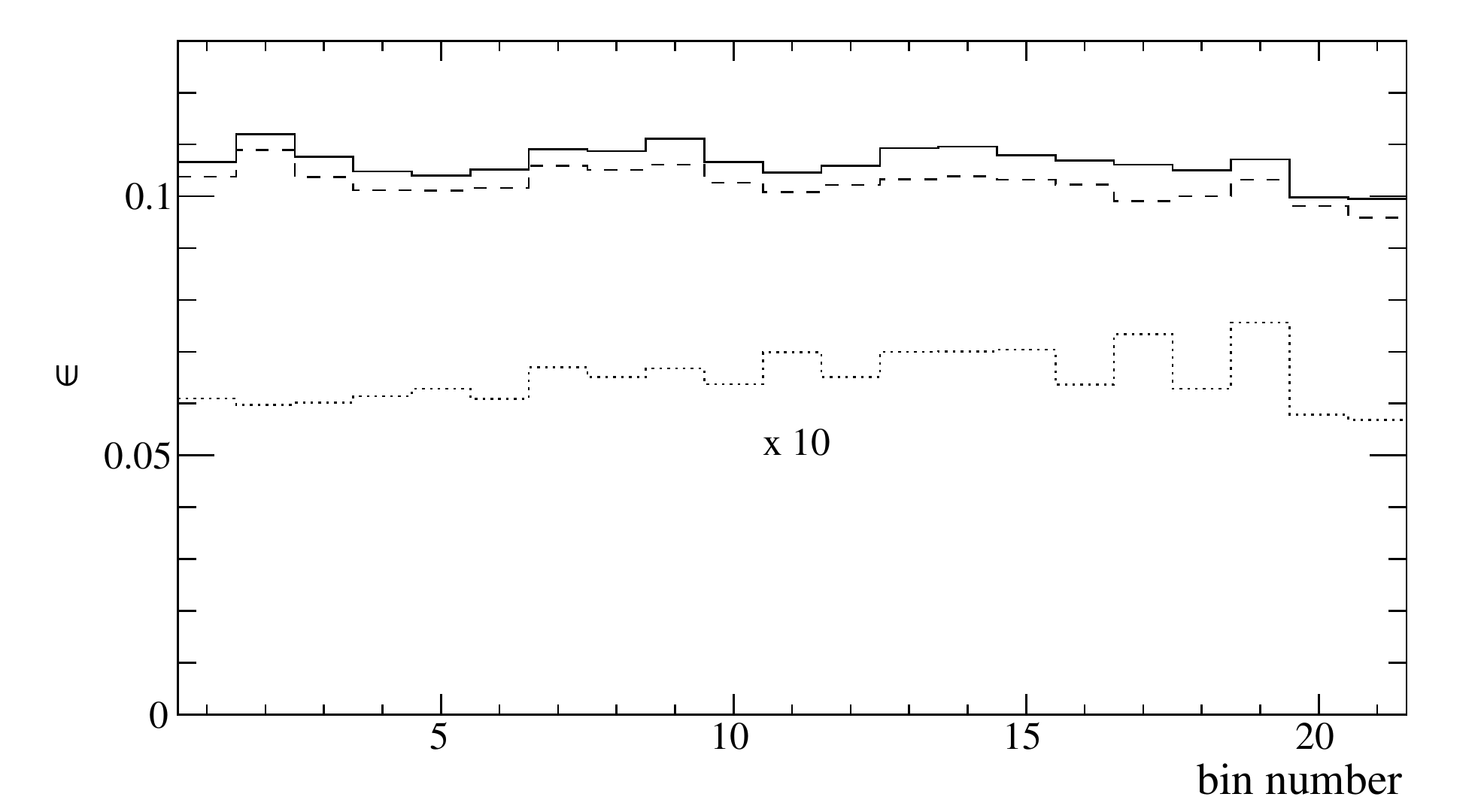}
\caption{The resulting efficiencies for each Dalitz plot bin for the three data sets. The solid line corresponds to set~A, the dashed line to set~B, and the dotted line is the acceptance for the $pp$ data set~C, which is multiplied by a factor of 10.}\label{fig:AccCorr}
\end{figure}

Since the $P$-wave distribution reproduces the general features of the $\omega\to\pi^+\pi^-\pi^0$ Dalitz plot very well and the deviations are expected to be small, the efficiency correction is obtained using signal simulation with the $P$-wave. The efficiency, $\epsilon_{i}$, is extracted using the ratio $\epsilon_{i}=N_{i}/N^{G}_{i}$. $N_{i}^{G}$ is the number of events with generated kinematic variables corresponding to bin $i$ in the Dalitz plot. $N_{i}$ is the content of the bin $i$ when the reconstructed values of the kinematic variables are used for events passing all analysis steps. The extracted efficiencies for the three data sets are shown in Fig.~\ref{fig:AccCorr}. For the $pd$ data sets the overall efficiency is 11\%, while for $pp$ it is 20 times lower. This low efficiency for the $pp$ data sample has the following well-understood causes. In most cases, the two fast proton tracks deposit only a fraction of their kinetic energy in the detector, leading to a lower precision of the kinetic energy determination and an asymmetric resolution function. The events from the tails will likely be rejected by the kinematic fit procedure. On the other hand for the $pd \rightarrow\mbox{}^{3}\mbox{He}\, \omega$ reaction, there is only one doubly charged $^3$He stopping in the detector. Another cause for the low efficiency is the larger centre-of-mass velocity in the $pp$ reaction, which decreases the average emission angle for decay particles, in particular for the charged pions. The pions will be more often emitted at angles below 30$^\circ$ and will therefore be rejected in the analysis procedure.

The Dalitz plot parameters ($\alpha$, $\beta$, $\ldots$) and normalisation factors for the three data sets ($\mathcal{N}_A$, $\mathcal{N}_B$, $\mathcal{N}_C$) are determined by minimising the following $\chi^2=\chi^2_A+\chi^2_B+\chi^2_C$ function, where 

\begin{equation}
\chi^2_A = \sum\limits_{i} \left ( \frac{\tilde{N}_{iA} - \mathcal{N}_A\cdot H_i(\alpha,\beta,\ldots)}
{\tilde{\sigma}_{iA}} \right )^2.\label{eq:DPFitFunc}
\end{equation}

$\tilde{N}_{i}$ and $\tilde{\sigma}_{i}$ are the efficiency corrected experimental Dalitz plot bin content and error, respectively. $H_{i}$ is given by an integral over bin $i$:
$H_i(\alpha,\beta,\ldots)=\int_i \mathcal{P}(Z,\phi)\mathcal{G}(Z,\phi) dZd\phi$. $\mathcal{P}(Z,\phi)$ is the $P$-wave phase space term given by Eq.~\eqref{eq:pwave}, calculated using the nominal mass of the $\omega$ meson of 782.65\,MeV, and $\mathcal{G}(Z,\phi)$ is given by Eq.~\eqref{eq:DPamplitude}.

The parametrisation procedure of the Dalitz plot is tested using $10^6$ signal events simulated with $P$-wave phase space only ({\it i.e.}\ ${\mathcal G}=1$) and without detector smearing. The extracted parameters are found to be consistent with zero and therefore the procedure does not introduce any bias at the present statistical accuracy.  

The three independent data sets and the Dalitz plot symmetries allow for detailed checks of the experimental efficiency and the background subtraction procedure since the background distributions and efficiencies are different in the corresponding bins. 

The method of background subtraction for the missing mass $\mu$ distributions is tested by preparing simulated distributions after full detector reconstruction, consisting of a sum of $\pi^+\pi^-\pi^0$ production background events and the $\omega$ signal generated using a $P$-wave phase space distribution. The background is obtained from the $\big\{a_0 + a_1\mu + a_2\mu^2+ a_3\mu^3 \big\}\times H_ {3\pi}(\mu)$ distributions with $a_i$ determined from the fits using Eq.~\eqref{eq:MMFitFunc} and by setting the average signal-to-background ratio to be the same as in the experimental data.
The generated $\mu$ distributions with the number of events similar as in the experiment are then subjected to the same background subtraction as the experimental data. The combined fit of the Dalitz plot parametrisation to the samples A and B with only the $\alpha$ parameter gives $\alpha = (10\pm 35)\cdot 10^{-3}$ and $\chi^2=36/39$. For set~C $\alpha = (25\pm 59)\cdot 10^{-3}$ and $\chi^2= 24 / 19$. Therefore the background subtraction procedure does not introduce any experimental bias.

\begin{figure}[t!]
\includegraphics*[width=\linewidth]{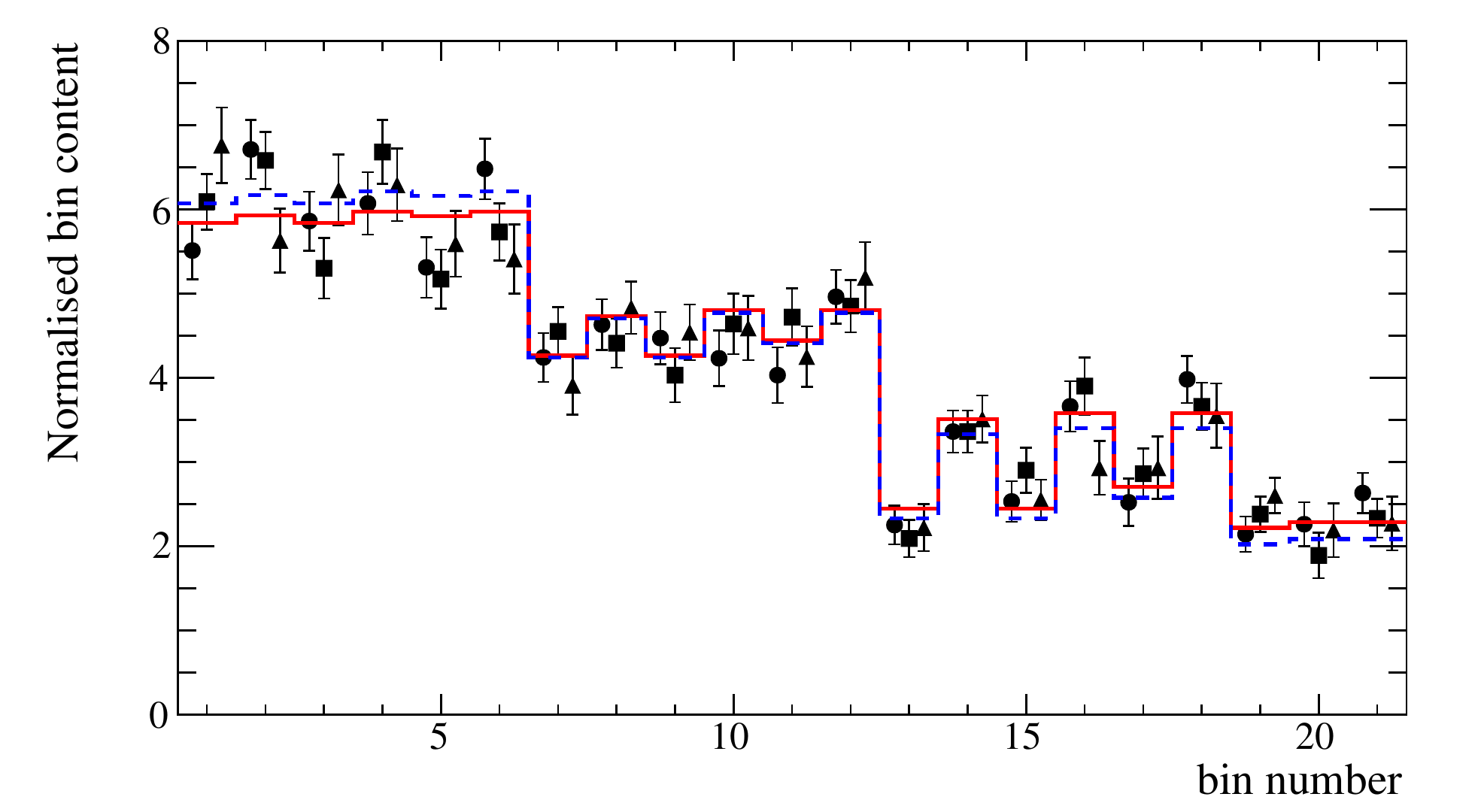}
\caption{The experimental Dalitz plot distribution after applying an efficiency correction. Circles correspond to set~A, squares to set~B, and triangles to set~C. The solid red line is the standard fit result (with $\alpha$ parameter), and the dashed line is $P$-wave only.}\label{fig:ExpDP}
\end{figure}
\begin{table}[t!]
\centering
\begin{tabular}{c|rrr}
\toprule
bin\#& set A& set B& set C\\
\midrule
1 & 5.51(34) & 6.09(33) & 6.76(45)\\
2 & 6.71(35) & 6.58(34) & 5.63(38)\\
3 & 5.86(35) & 5.30(36) & 6.23(42)\\
4 & 6.07(37) & 6.68(38) & 6.29(43)\\
5 & 5.31(36) & 5.17(35) & 5.59(39)\\
6 & 6.48(36) & 5.73(34) & 5.41(41)\\
7 & 4.24(29) & 4.55(29) & 3.91(35)\\
8 & 4.63(30) & 4.41(29) & 4.83(31)\\
9 & 4.47(31) & 4.03(32) & 4.54(33)\\
10 & 4.23(33) & 4.64(36) & 4.59(38)\\
11 & 4.03(33) & 4.72(34) & 4.25(36)\\
12 & 4.96(32) & 4.85(31) & 5.19(42)\\
13 & 2.25(23) & 2.09(22) & 2.22(28)\\
14 & 3.36(25) & 3.36(25) & 3.51(28)\\
15 & 2.53(24) & 2.90(27) & 2.55(24)\\
16 & 3.66(30) & 3.90(34) & 2.93(32)\\
17 & 2.52(28) & 2.86(30) & 2.93(37)\\
18 & 3.98(28) & 3.66(28) & 3.55(38)\\
19 & 2.14(21) & 2.38(21) & 2.60(21)\\
20 & 2.26(26) & 1.89(27) & 2.19(32)\\
21 & 2.63(24) & 2.33(23) & 2.27(32)\\
\bottomrule
\end{tabular}
\caption{Dalitz plot bin content for the three data sets. The relative normalisation between the sets is based on the normalisation factors ($\mathcal{N}_A$, $\mathcal{N}_B$, $\mathcal{N}_C$) obtained from individual fits of the $\alpha$ parameter to the three data sets. The overall normalisation factor is arbitrary.} 
\label{tab:DPdata}
\end{table}

One can also study the bias and accuracy of the efficiency determination by considering $X$- or $Y$-dependent corrections for the efficiency: $\epsilon_{i}\to\epsilon_i\cdot(1+\xi_AX)$ or $\epsilon_{i}\to\epsilon_i\cdot(1+\zeta_AY)$, where $\xi_{A,\ldots}$, $\zeta_{A,\ldots}$ are single parameters for each data set. Fits to separate data sets show that all $\zeta$ coefficients are consistent with zero and do not change the value of the $\chi^2$. On the other hand, $\xi_B$ and $\xi_C$ were found to significantly deviate from zero, although with opposite signs. Applying these two corrections to the efficiency before a fit of the Dalitz plot parametrisation yields a significantly reduced $\chi^2$ value. However, the determined values of the Dalitz parameters are not affected, \textit{e.g.}\ $\alpha=(147\pm 35)\cdot 10^{-3}$ and $\alpha=(147\pm 36)\cdot 10^{-3}$ without and with correction, respectively.  This comes from the fact that the fitted parametrisation is preserving isospin symmetry. In conclusion, we apply the $X$-dependent corrections to the efficiency corrections of data sets~B and C, as it ensures the anticipated charge symmetry of the Dalitz plot and leads to a decrease of the $\chi^2$ as well as the correct statistical significance when fitting the Dalitz parameters. The resulting efficiency corrected and normalised Dalitz plot bin contents are provided in Table~\ref{tab:DPdata}.

\begin{table}
\centering
\renewcommand{\arraystretch}{1.2}
\begin{tabular}{crrr}
\toprule
{Data set}  & $\alpha\times 10^3$& $\beta\times 10^3$ &  $\chi^2/\text{d.o.f.}$\\
\midrule
\multirow{3}{*}{$A$} & \multicolumn{1}{c}{--} & \multicolumn{1}{c}{--} &  28.7 / 20 \\
                     & 142(59) & \multicolumn{1}{c}{--} & 22.2 / 19 \\
 & 102(66) & 109(87) &  20.7 / 18  \\
\midrule
\multirow{3}{*}{$B$} & \multicolumn{1}{c}{--} & \multicolumn{1}{c}{--}
 &  35.4 / 20 \\
 & 146(59) & \multicolumn{1}{c}{--} &  28.5 / 19 \\
 & 154(69) & $-21$(92) & 28.5 / 18\\
\midrule
\multirow{3}{*}{$C$} & \multicolumn{1}{c}{--} & \multicolumn{1}{c}{--} 
 &  26.5 / 20\\
 & 154(69) & \multicolumn{1}{c}{--} &  20.8 / 19\\
 & 149(78) &  14(102) & 20.8 / 18\\
\bottomrule
\end{tabular}
\caption{The resulting Dalitz plot parameters after a individual fits to the three data sets, where at most two parameters were used in the fit.}\label{tab:DPdatafitparInd}
\renewcommand{\arraystretch}{1.0}
\end{table}

\begin{table}
\centering
\renewcommand{\arraystretch}{1.2}
\begin{tabular}{lrrr}
\toprule
 & $\alpha\times 10^3$& $\beta\times 10^3$&  $\chi^2/\text{d.o.f.}$\\
\midrule
& \multicolumn{1}{c}{--} & \multicolumn{1}{c}{--} & 90.6 / 60 \\
& 147(36) & \multicolumn{1}{c}{--} & 71.5 / 59\\
& 133(41) & 37(54) & 71.0 / 58 \\
\bottomrule
\end{tabular}
\caption{Dalitz plot parameters from simultaneous fits to the three data sets, where at most two parameters were used in the fit.}\label{tab:DPdatafitpar}
\renewcommand{\arraystretch}{1.0}
\end{table}

The extracted Dalitz plot parameters and goodness of fit for each data set separately are reported in Table~\ref{tab:DPdatafitparInd}. There is a significant decrease of the $\chi^2$ value when including the $\alpha$ parameter and the results from the three data sets are consistent. The results of the fits for all data sets combined are given in Table~\ref{tab:DPdatafitpar}. The $p$-value significantly improves after including the $\alpha$ parameter, while inclusion of an additional parameter does not improve the $p$-value any further. The efficiency corrected Dalitz plots for the three data sets are shown in Fig.~\ref{fig:ExpDP}, where they are compared to the $P$-wave  distribution as well as the fit with the $\alpha$ parameter. We consider the result with the $\alpha$ parameter our main finding. The difference between the results from the first and second fits in Table~\ref{tab:DPdatafitpar} indicates the onset of dynamics in the reaction on top of the $P$-wave phase space distribution. This follows the expected behaviour of an increase towards the edges of phase space due to the attractive $\pi\pi$ final-state interaction (approaching the $\rho$ resonance), yielding a positive value for the $\alpha$ parameter.

\begin{figure}
\includegraphics*[width=\linewidth]{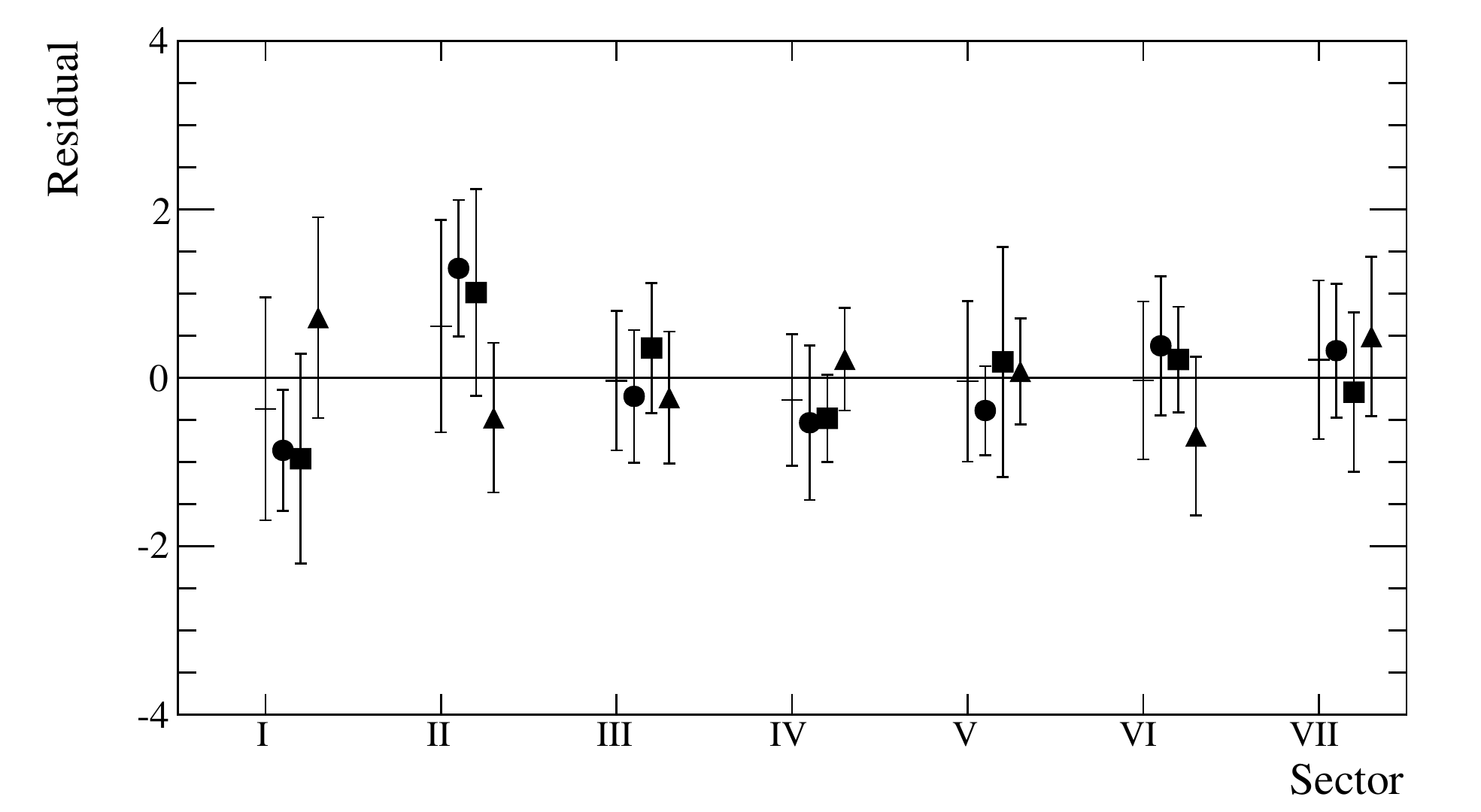}
\caption{Arithmetic averages and root mean square of the normalised residuals in separate Dalitz plot sectors for all data sets (crosses) and for the separate data sets (A -- circles, B -- squares, C -- triangles) for the standard fit.
}\label{fig:Resid}
\end{figure}
Figure~\ref{fig:Resid} shows a test of data consistency for the same Dalitz plot sectors, marked with Roman numerals in Fig.~\ref{fig:DPbins}. Arithmetical averages of the normalised residuals with respect to the $\alpha$ parameter fit from bins corresponding to the same sectors are calculated for the separate data sets and for all three data sets combined. The error bars correspond to the calculated root mean square values, which are expected to be 1 for a random data sample with correctly estimated uncertainties.

The conclusion of the checks for the systematic effects is that the accuracy is dominated by statistic uncertainty.

\section{Summary and discussion}

\begin{figure}
\includegraphics*[width=\linewidth]{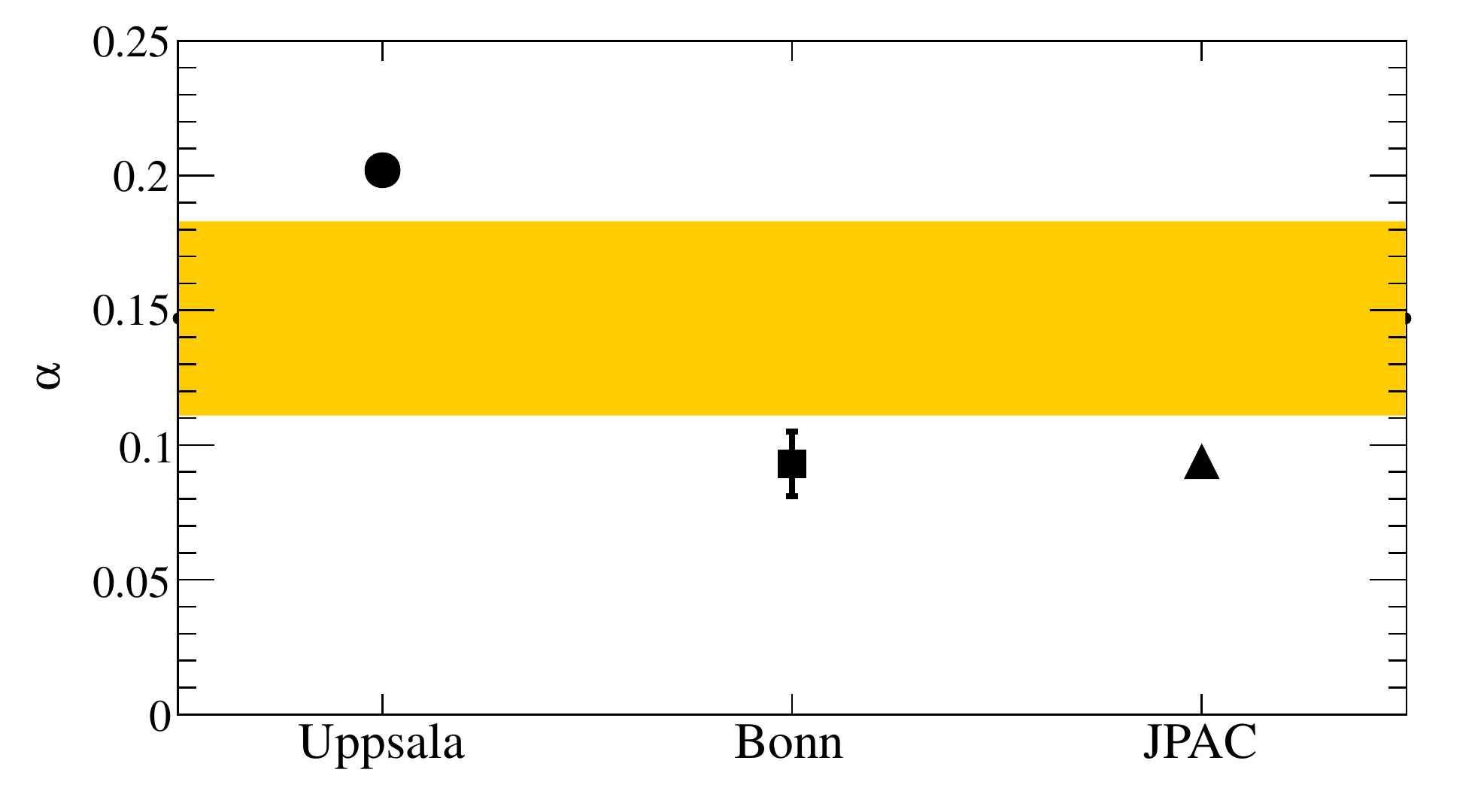}
\caption{Comparison of our result for the $\alpha$ parameter 
(shaded area) with the three theoretical predictions~\cite{Terschlusen:2013iqa,Niecknig:2012sj,Danilkin:2014cra}.
}\label{fig:DPthexp}
\end{figure}

For the first time a deviation from a pure $P$-wave distribution in $\omega\to\pi^+\pi^-\pi^0$ is observed and quantified by the determination of the parameter $\alpha = (147\pm 36)\cdot 10^{-3}$, {\it i.e.}\ a positive value with 4.1$\sigma$ significance. Figure~\ref{fig:DPthexp} compares the experimental result of the $\alpha$ parameter to theoretical predictions. The systematic effects were studied by comparing three data sets using two production reactions, which differ significantly in resolution and acceptance. The chosen $Z$, $\phi$ parametrisation together with isospin symmetry allows for more tests of systematic effects, and the precision of the result is found to be dominated by the statistical uncertainty.

\section*{Acknowledgments}
This work was supported in part by the EU Integrated Infrastructure Initiative HadronPhysics Project under contract number RII3--CT--2004--506078; by the European Commission under the 7th Framework Programme through the
Research Infrastructures action of the Capacities Programme, Call: FP7--INFRASTRUCTURES--2008-1, Grant Agreement N.\ 227431; by the Polish National Science Centre through the grants DEC--2013/11/N/ST2/04152, 2011/01/B/ST2/00431, 2011/03/B/ST2/01847, and the Foundation for Polish Science (MPD), co--financed by the European Union within the European Regional Development Fund. We gratefully acknowledge the support given by the Swedish Research Council, the Knut and Alice Wallenberg Foundation, and the Forschungszentrum J\"ulich FFE Funding Program. This  work is based on the PhD theses  of Lena Heijkenskj\"old and Siddhesh Sawant. 

\raggedright
\bibliographystyle{./model1a-num-names}
\bibliography{./omega3pi}

\end{document}